\documentclass[letter,longauth,traditabstract,a4paper]{aa}    
\usepackage[usenames]{color}
\usepackage{graphicx}      
\usepackage{txfonts}
\usepackage{natbib}
\begin{document}
\title{Herschel-PACS observation of the 10 Myr old T~Tauri disk
  \object{TW~Hya} \thanks{Herschel is an ESA space observatory with
    science instruments provided by Principal Investigator
    consortia. It is open for proposals for observing time from the
    worldwide astronomical community.}}

   \subtitle{Constraining the disk gas mass}

   \author{
     W.-F. Thi\inst{1, 2}, 
     G. Mathews\inst{3}, 
     F. M\'enard\inst{2}, 
     P. Woitke\inst{4, 5, 1},
     G. Meeus\inst{6},
     P. Riviere-Marichalar\inst{7},
     C. Pinte\inst{2,8}, 
     C. D. Howard\inst{9}, 
     A. Roberge\inst{10}, 
     G. Sandell\inst{9},
     I. Pascucci\inst{11}, 
     B. Riaz\inst{11},
     C. A. Grady\inst{12}, 
     W.~R.~F. Dent\inst{13}, 
     I. Kamp\inst{14},
     G. Duch\^ene\inst{2, 15}, 
     J.-C. Augereau\inst{2},
     E. Pantin\inst{16},
     B. Vandenbussche\inst{17},
     I. Tilling\inst{1},
     J. P. Williams\inst{3}, 
     C. Eiroa\inst{6}, 
     D. Barrado\inst{18, 7}, 
     J. M. Alacid\inst{19, 20}, 
     S. Andrews\inst{21},   
     D.R. Ardila\inst{22},
     G. Aresu\inst{14}, 
     S. Brittain\inst{23},
     D.~R. Ciardi\inst{24}, 
     W. Danchi\inst{25}, 
     D. Fedele\inst{26, 27, 28}, 
     I. de Gregorio-Monsalvo\inst{13}, 
     A. Heras\inst{29},
     N. Huelamo\inst{4}, 
     A. Krivov\inst{30}, 
     J. Lebreton\inst{2},
     R. Liseau\inst{31}, 
     C. Martin-Zaidi\inst{2}, 
     I. Mendigut\'ia\inst{4}, 
     B. Montesinos\inst{4}, 
     A. Mora\inst{31},
     M. Morales-Calderon\inst{32}, 
     H. Nomura\inst{33}, 
     N. Phillips\inst{1}, 
     L. Podio\inst{14},
     D.~R. Poelman\inst{5}, 
     S. Ramsay\inst{34}, 
     K. Rice\inst{1}, 
     E. Solano\inst{19, 20}, 
     H. Walker\inst{35},
     G.~J. White\inst{36, 35}, 
     G. Wright\inst{4}}

\institute{
SUPA, Institute for Astronomy, University of Edinburgh, Royal Observatory Edinburgh, UK 
\email{wfdt@roe.ac.uk} 
\and 
Universit\'e Joseph-Fourier – Grenoble 1/CNRS, Laboratoire d’Astrophysique de Grenoble (LAOG) UMR 5571, BP 53, 38041 Grenoble Cedex 09, France
\and 
Institute for Astronomy, University of Hawaii at Manoa, Honolulu, HI 96822, USA
\and 
UK Astronomy Technology Centre, Royal Observatory, Edinburgh, Blackford Hill, Edinburgh EH9 3HJ, UK
\and 
School of Physics \& Astronomy, University of St.~Andrews, North Haugh, St.~Andrews KY16 9SS, UK
\and 
Dep. de F\'isica Te\'orica, Fac. de Ciencias, UAM Campus Cantoblanco, 28049 Madrid, Spain
\and 
LAEX, Depto. Astrof{\'i}sica, Centro de Astrobiolog{\'i}a (INTA-CSIC), P.O. Box 78, E-28691 Villanueva de la Ca\~nada, Spain
\and 
School of Physics, University of Exeter, Stocker Road, Exeter EX4 4QL, United Kingdom 
\and 
SOFIA-USRA, NASA Ames Research Center, Mailstop 211-3 Moffett Field CA 94035 USA
\and 
Exoplanets and Stellar Astrophysics Lab, NASA Goddard Space Flight Center, Code 667, Greenbelt, MD, 20771, USA 
\and 
Space Telescope Science Institute, 3700 San Martin Drive, Baltimore, MD 21218, USA
\and 
Eureka Scientific and Exoplanets and Stellar Astrophysics Lab, NASA Goddard Space Flight Center, Code 667, Greenbelt, MD, 20771, USA
\and 
ESO-ALMA, Avda Apoquindo 3846, Piso 19, Edificio Alsacia, Las Condes, Santiago, Chile
\and 
Kapteyn Astronomical Institute, P.O. Box 800, 9700 AV Groningen, The
Netherlands \and 
Astronomy Department, University of California, Berkeley CA 94720-3411
USA \and 
CEA/IRFU/SAp, AIM UMR 7158, 91191 Gif-sur-Yvette, France
\and 
Instituut voor Sterrenkunde, KU Leuven, Celestijnenlaan 200D, 3001
Leuven, Belgium \and 
Calar Alto Observatory, Centro Astron\'omico Hispano-Alem\'an
C/Jes\'us Durb\'an Rem\'on, 2-2, 04004 Almer\'{\i}a, Spain
\and 
Unidad de Archivo de Datos, Depto. Astrof{\'i}sica, Centro de
Astrobiolog{\'i}a (INTA-CSIC), P.O. Box 78, E-28691 Villanueva de la
Ca\~nada, Spain \and 
Spanish Virtual Observatory \and 
Harvard-Smithsonian Center for Astrophysics, 60 Garden St., Cambridge,
MA, USA \and 
NASA Herschel Science Center, California Institute of Technology,
Pasadena, CA, USA.  \and 
Clemson University, Clemson, SC, USA
 \and 
NASA Exoplanet Science Institute/Caltech 770 South Wilson Avenue, Mail
Code: 100-22, Pasadena, CA USA 91125 \and 
NASA Goddard Space Flight Center, Exoplanets \& Stellar Astrophysics,
Code 667, Greenbelt, MD 20771, USA \and 
Departamento de Fisica Teórica, Facultad de Ciencias, Universidad
Autónomade Madrid, Cantoblanco, 28049 Madrid, Spain \and 
Max Planck Institut f{\"u}r Astronomie, K{\"o}nigstuhl 17, 69117
Heidelberg, Germany \and 
Johns Hopkins University Dept. of Physics and Astronomy, 3701 San
Martin drive Baltimore, MD 21210 USA \and 
Research and Scientific Support Department-ESA/ESTEC, PO Box 299, 2200
AG Noordwijk, The Netherlands \and 
Astrophysikalisches Institut und Universit{\"a}tssternwarte,
Friedrich-Schiller-Universit{\"a}t, Schillerg{\"a}{\ss}chen 2-3, 07745
Jena, Germany \and 
Department of Radio and Space Science, Chalmers University of
Technology, Onsala Space Observatory, 439 92 Onsala, Sweden
\and 
ESA-ESAC Gaia SOC, P.O. Box 78. E-28691 Villanueva de la Ca\~{n}ada,
Madrid, Spain \and 
Department of Astronomy, Graduate School of Science, Kyoto University,
Kyoto 606-8502,Japan \and 
European Southern Observatory, Karl-Schwarzschild-Strasse, 2, 85748
Garching bei M\"unchen, Germany.  \and 
The Rutherford Appleton Laboratory, Chilton, Didcot, OX11 OQL, UK
\and 
Department of Physics \& Astronomy, The Open University, Milton Keynes
MK7 6AA, UK and The Rutherford Appleton Laboratory, Chilton, Didcot,
OX11 OQL, UK }
\authorrunning{Thi et al.}
\titlerunning{Herschel observations of TW~Hya}            

   \date{Received 31 March 2010; accepted 28 April 2010}

   
   \abstract{Planets are formed in disks around young stars.  With an
     age of $\sim$~10\,Myr, \object{TW~Hya} is one of the nearest
     T~Tauri stars that is still surrounded by a relatively massive
     disk. In addition a large number of molecules has been found in
     the \object{TW~Hya} disk, making \object{TW~Hya} the perfect test
     case in a large survey of disks with {\it Herschel}--{\it PACS}
     to directly study their gaseous component. We aim to constrain
     the gas and dust mass of the circumstellar disk around
     \object{TW~Hya}. We observed the fine-structure lines of
     [\ion{O}{I}] and [\ion{C}{II}] as part of the Open-time large
     program {\it GASPS}. We complement this with continuum data and
     ground-based $^{12}$ CO 3--2 and $^{13}$CO 3--2 observations.  We
     simultaneously model the continuum and the line fluxes with the
     3D Monte-Carlo code {\it MCFOST} and the thermo-chemical code
     {\it ProDiMo} to derive the gas and dust masses. We detect the
     [\ion{O}{I}] line at 63 $\mu$m.  The other lines that were
     observed, [\ion{O}{I}] at 145~$\mu$m and [\ion{C}{II}] at
     157~$\mu$m, are not detected. No extended emission has been
     found. Preliminary modeling of the photometric and line data
     assuming [$^{12}$CO]/[$^{13}$CO]=69 suggests a dust mass for
     grains with radius $<$ 1~mm of $\sim$ 1.9 $\times$ 10$^{-4}$
     M$_{\odot}$ (total solid mass of 3 $\times$ 10$^{-3}$
     M$_{\odot}$) and a gas mass of (0.5--5) $\times$ 10$^{-3}$
     M$_{\odot}$. The gas-to-dust mass may be lower than the standard
     interstellar value of 100.}

   \keywords{Circumstellar disks}

   \maketitle
%

\section{Introduction}

Planets are formed in the disks that surround a large fraction of
T~Tauri stars. Knowledge of the gas mass available at different disk
ages is essential to constrain giant planet formation models. Most
studies estimate the dust mass from millimeter continuum emission and
assume the gas mass is a factor of 100 times larger.  This conversion
factor has been calibrated for the interstellar medium but is likely
not valid for disks, especially those that are evolving toward debris
disks or where most of the gas has accreted onto the planetary
atmosphere. Disk gas mass estimates derived from observations of
$^{12}$CO and optically thinner $^{13}$CO emission are at least a
factor of 10 lower than the mass derived from dust observations assuming
the interstellar medium conversion factor.  The discrepancy has been
ascribed to CO photodissociation at disk atmosphere and freeze-out
onto cold dust grains in the disk midplane
\citep[e.g.,][]{Qi2004ApJ...616L..11Q,Thi2001ApJ...561.1074T}. An
alternative explanation is that the CO abundance is not different and
the gas in disks has been depleted.

The {\it PACS} instrument \citep{Poglistch2010} on-board the {\it
  Herschel Space Telescope} \citep{Pilbratt2010} makes it possible to
observe lines from species that result from the photodissociation of
CO (atomic oxygen and singly ionized carbon). With observations of all
the major gas-phase carbon and oxygen-bearing species, we can more
precisely constrain the disk gas mass.

At a distance of $\sim$~56~pc \citep{Wichmann1998MNRAS.301L..39W},
\object{TW~Hya} is one of the nearest classical T~Tauri stars with an
estimated age of 10~Myr \citep{Barrado2006A&A...459..511B}.  Its
proximity allows us to attain an order of magnitude higher mass
sensitivity than objects in the Taurus molecular cloud.  Fits to the
spectral energy distribution (SED) provide an estimate of the gas disk
mass of 6 $\times$ 10$^{-2}$ M$_{\odot}$ after applying a conversion
factor of $\sim$~75 \citep{Calvet2002ApJ...568.1008C}. This large disk
mass at this advanced age is surprising as the median disk lifetime is
only 2-3 Myr \citep{Haisch2001ApJ...553L.153H}. \object{TW~Hya} is
considered a transition object with an optically thin inner cavity and
an optically thick outer disk
\citep{Calvet2002ApJ...568.1008C,Ratzka2007A&A...471..173R}. The fit
to the SED also suggests that grains have grown to at least
$\sim$~1~cm.  

The star \object{TW~Hya} was observed as a Science Demonstration
Project object and is part of the {\it Herschel-GASPS} program
\citep{Dent2010}. {\it Herschel} observations of the disk around the
Herbig~Ae star \object{HD169142} are presented by
\citet{Meeus2010}. In this letter we use fine-structure lines in
addition to continuum data and CO (sub)millimeter lines to directly
constrain the gas mass and compare it to the dust mass derived from
fits to the SED.

\begin{figure}[!ht]
\centering
{\includegraphics[angle=90,width=8cm]{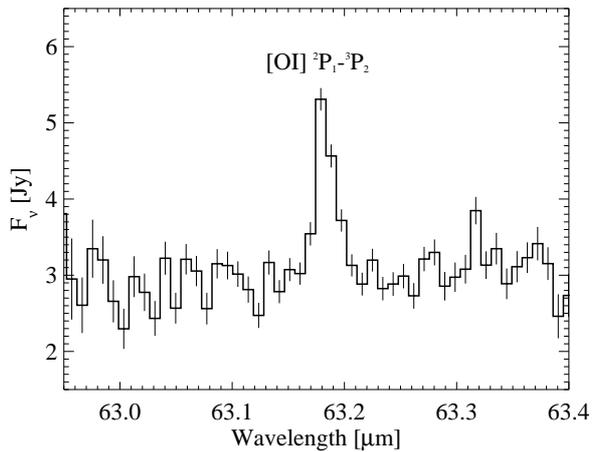}}
\caption{{\it Herschel-PACS} spectrum centred around the OI 63 $\mu$m
  line on the upper-left panel.}
  \label{fig_OI63_herschel}          
\end{figure}   

\begin{table}
\centering
\small
\caption{Lines observed by {\it Herschel-PACS}. The errors and upper limits are 3~$\sigma$. The calibration error adds an extra $\sim$~40\% uncertainty. The CO data also have uncertainties of 30\%.}
\label{table_results} 
\begin{tabular}{lllll}   
\hline 
\noalign{\smallskip}   
\multicolumn{1}{c}{Line} & \multicolumn{1}{c}{Cont. flux} & \multicolumn{1}{c}{Obs.}&\multicolumn{1}{c}{GH08}&\multicolumn{1}{c}{M08}\\
           &  \multicolumn{1}{c}{(Jy)}& \multicolumn{3}{c}{(10$^{-18}$ W m$^{-2}$)}\\  

\noalign{\smallskip} 
\hline
\noalign{\smallskip} 
O{\sc I}\ $^3$P$_1 \rightarrow ^3$P$_2$ & 2.99~$\pm$~0.14 & \phantom{$<$}36.5~$\pm$~12.1 & 124-161 & 412\\
O{\sc I}\ $^3$P$_0 \rightarrow ^3$P$_1$ & 7.00~$\pm$~0.05 &       $<$\  \ 5.5 & 25-41 & 11\\
C{\sc II}\ $^2$P$_{3/2} \rightarrow ^2$P$_{1/2}$ & 8.79~$\pm$~0.08  &       $<$\  \ 6.6& 0.8-12 & 0.06\\
CO 3--2 & \multicolumn{1}{c}{n.a.} & \phantom{$<$}\ \ 0.43 & 0.3-0.6 & n.a.\\
$^{13}$CO 3--2 & \multicolumn{1}{c}{n.a.} & \phantom{$<$}\ \ 4.4 $\times$ 10$^{-2}$& n.a.& n.a.\\     
\noalign{\smallskip} 
\hline
\end{tabular}
\end{table}
\normalsize 

\section{Observations and results}

We obtained photometry in the ``blue'' (70~$\mu$m) and ``red''
(160~$\mu$m) band of the {\it PACS} camera by doing mini scan maps
with a scan speed of $20''$ and a scan length of $2'$ (obsid
1342187342). The total duration of this map was 731\,sec, with an
on-source time of 146\,seconds. The results are 3.90 $\pm$ 0.02 Jy and
7.38 $\pm$ 0.04 in the blue and red band respectively and have an
absolute accuracy estimated to be 5\% for the blue channel and 10\%
for the the red channel.  These values agree very well with the
observed IRAS flux densities and also with the continuum flux 
densities measured with the {\it PACS} spectrometer
(Table~\ref{table_results}). We also used the {\it PACS} spectrometer
to target the [OI] line at 63 $\mu$m in line scan mode, and the [OI]
and [CII] lines at 145 and 158 $\mu$m, respectively in range scan mode
(obsid 1342187127 PacsLineSpec and obsid 1342187238
PacsRangeSpec). Only the [OI] line at 63 $\mu$m was detected and we
report upper limits for the other two lines; see
Table~\ref{table_results}.  The absolute accuracy of {\it PACS}
spectroscopy is currently estimated to be about 40\%, but is expected
to improve in the future.  Figure~\ref{fig_OI63_herschel} shows the
spectrum centered at the position of the O{\sc I} line at 63 $\mu$m of
the central pixel.

\begin{center}
\begin{table}
  \caption{Disk parameters for the modeling.}\label{disk_parameters}
		\begin{tabular}{llcc}
                  \hline
 \noalign{\smallskip}   
                 \multicolumn{4}{c}{Fixed parameters}\\
\noalign{\smallskip}   

                  & & Inner cavity & Outer ring\\
                  Stellar mass & $M_*$(M$_\odot$) &  \multicolumn{2}{c}{0.6} \\ 
                  Stellar luminosity &$L_*$(L$_\odot$)  &  \multicolumn{2}{c}{0.23} \\ 
                  Effective temperature & $T_{\mathrm {eff}}$(K) & \multicolumn{2}{c}{4000}\\
                  Solid material mass density & $\rho_{\mathrm{dust}}$(g cm$^{-3}$) & \multicolumn{2}{c}{3.5} \\
                  Inner radius                      & $R_{\mathrm{in}}$(AU) & 0.25 & 4 \\
                  Outer radius                      & $R_{\mathrm{out}}$(AU) & 4 & 196\\
                   ISM UV field   & $\chi$          &  \multicolumn{2}{c}{1.0}\\
                  $\alpha$ viscosity parameter      & $\alpha$           & \multicolumn{2}{c}{0.0}\\
                  Turbulent velocity & $v_{\mathrm{turb}}$(km s$^{-1}$) & \multicolumn{2}{c}{0.05}\\
                  Disk inclination & $i$ & \multicolumn{2}{c}{7}\\
             CO isotopologue ratio & [$^{12}$CO]/[$^{13}$CO] & \multicolumn{2}{c}{69}\\
\noalign{\smallskip}   
                  \hline
\noalign{\smallskip}   
                  \multicolumn{4}{c}{{\it MCFOST} best fit parameters}\\
\noalign{\smallskip}   
                  Column density index  & $\epsilon$ & \multicolumn{2}{c}{1} \\
                  Reference scale height            & $H_0$(AU)           & 2.0 & 10.0\\
                  Reference radius                  &                    & 100 & 100\\
                  Flaring index                     & $\gamma$            & 0.6   & 1.12\\
                  Minimum grain size       & $a_{\mathrm{min}}$($\mu$m) & \multicolumn{2}{c}{3 $\times$ 10$^{-2}$}\\
                  Maximum grain size       & $a_{\mathrm{max}}$(cm) & \multicolumn{2}{c}{10}\\
                  Dust size distribution index  & $p$               & \multicolumn{2}{c}{3.4}\\
                  Dust mass ($a<$1 mm) & $M_{\mathrm{dust}}$(M$_{\odot}$) & 1.2 $\times$ 10$^{-9}$ & 1.9 $\times$ 10$^{-4}$\\
                  Solid mass           &$M_{\mathrm{solid}}$(M$_{\odot}$) & 2.0 $\times$ 10$^{-8}$  & 3.0 $\times$ 10$^{-3}$\\
\noalign{\smallskip}   
                  \hline
 \noalign{\smallskip}   
                  \multicolumn{4}{c}{{\it ProDiMo} parameter range}\\
\noalign{\smallskip}   
                  Disk gas mass                     & $M_{gas}$($M_{\odot}$) &  \multicolumn{2}{c}{3 $\times$ 10$^{-4}$--0.3}\\
                  UV excess                         & $F_{\mathrm{UV}}$ & \multicolumn{2}{c}{0.018}\\
                  Fraction of PAHs w.r.t. ISM  & $f_{\rm PAH}$      & \multicolumn{2}{c}{0.01, 0.1}\\
                  Cosmic ray flux                  & $\zeta$(s$^{-1}$)           & \multicolumn{2}{c}{(1.7--17) $\times$ 10$^{-17}$}\\
                  \hline
                \end{tabular}
\end{table}
\end{center}
\section{Modeling and discussion}

As there is no evidence for an outflow from \object{TW~Hya}, we assume
that all the fluxes arise from the circumstellar disk (see also the
discussion in \citealt{Mathews2010}). The interpretation of the
observations with the {\it DENT} grid of models is detailed in
\citet{Pinte2010}. We performed a more detailed analysis here.

\begin{figure}  
\centering
\resizebox{\hsize}{!}{\includegraphics[angle=0]{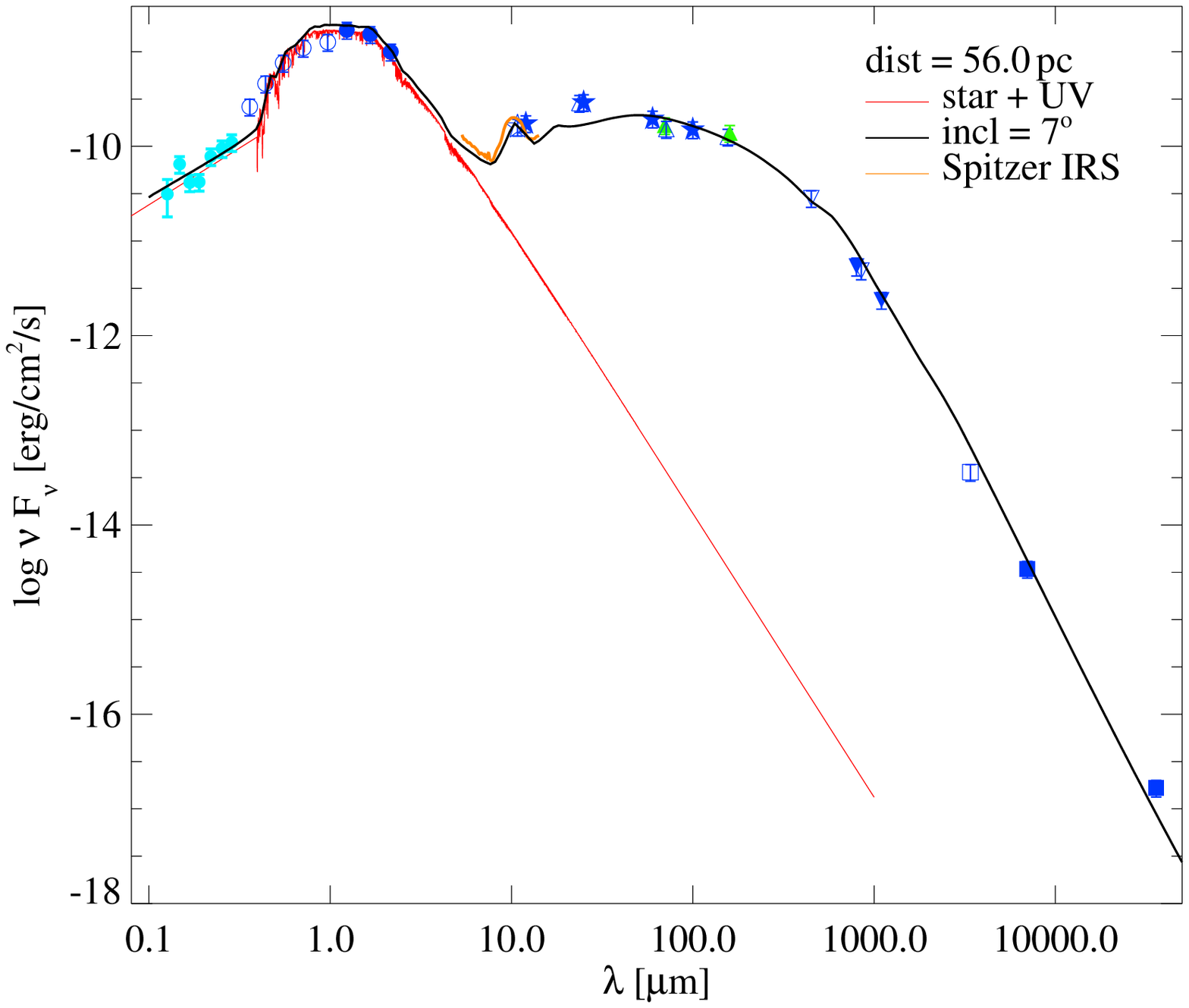}}
\caption{Fit to the SED generated by {\it ProDiMo} using the
  parameters from {\it MCFOST}.  The input {\it Phoenix} stellar
  spectrum plotted in red is from
  \citet{Brott2005ESASP.576..565B}. IUE (UV) data are from
  \citet{Valenti2003ApJS..147..305V}. The {\it 2MASS} J,H, K, {\it
    IRAS}, and {\it Spitzer-MIPS} photometry are archival data. The
  {\it Spitzer-IRS} spectrum is published by
  \citet{Ratzka2007A&A...471..173R}. The {\it Herschel-PACS} data are
  plotted in filled green triangles. The average {\it UBVRI}
  photometric points are published by
  \citet{Rucinski1983A&A...121..217R}. The 800 $\mu$m and 1.1mm data
  points (inverted filled blue triangles) are taken from
  \citet{Weintraub1989ApJ...340L..69W}. The 3.4mm point (open blue
  square) is from \citet{Wilner2003ApJ...596..597W} while the 7mm and
  3.6 cm points (filled blue square) are from
  \citet{Wilner2000ApJ...534L.101W}.}
  \label{fig_SED_ProDiMo_fit}  
\end{figure}

We first augmented the {\it Herschel} photometric data with continuum
measurements from the literature.  We also retrieved and reduced
archival {\it SCUBA} data for {\object TW Hya} obtained during two
nights with very good sub-millimeter transmission
($F_\nu$(450~$\mu$m)= 4.25 $\pm$ 0.85 Jy and $F_\nu$(850~$\mu$m)= 1.38
$\pm$ 0.14 Jy). The disk around \object{TW~Hya} has an internal cavity
from up to 4 AU where the gas and dust density are very low. Most of
the mass is located in the external ring. The inner
($R_{\mathrm{in}}$) and outer radius ($R_{\mathrm{out}}$) of the
external ring are well constrained by imaging studies and are fixed at
4 AU and 200 AU respectively
\citep{Roberge2005ApJ...622.1171R,Qi2004ApJ...616L..11Q,Hughes2007ApJ...664..536H}. We
fitted the SED with the 3D Monte-Carlo radiative transfer code {\it
  MCFOST} \citep{Pinte2006A&A...459..797P}. We chose to restrict to a
parametric disk model for this letter. The disk has a radial density
profile with index $\epsilon$. The flaring is characterized by an
opening angle $H_0$ at a given radius $R_{\mathrm{ref}}$ and a flaring
index $\gamma$ so that the gas scale-height is given by
$H=H_0(R/R_{\mathrm{ref}})^{\gamma}$. The low continuum flux in the
30--100 $\mu$m region suggests that the outer disk flaring is weak.
Amorphous olivine grains were used \citep{Dorschner95} with a
power-law size-distribution defined by a minimum radius
$a_{\mathrm{min}}$, maximum radius $a_{\mathrm{max}}$, and power-law
index $p$. The dust size-distribution and mass are well constrained by
the continuum emission at long wavelengths. The fit to the
long-wavelength photometric points including the new {\it
  Herschel-PACS} data is shown in Fig.~\ref{fig_SED_ProDiMo_fit} and
the disk parameters constrained by the fit are listed in
Table~\ref{disk_parameters}.  The inferred dust mass in grains with
radius $<$~1~mm is $M_{\mathrm{dust}}$=1.9 $\times$ 10$^{-4}$
M$_{\odot}$ and the total mass in solids (pebbles) up to
$a_{\mathrm{max}}=10$\,cm is $M_{\mathrm{solid}}$=3 $\times$ 10$^{-3}$
M$_{\odot}$. However, the fit fails to account for the flux at
$\sim$25~$\mu$m, which may stem from our assumption of a unique
temperature for grains of all sizes. The flux around 20-30 $\mu$m is
strongly inclination-dependent because we adopted a sharp density
change between the inner cavity and the outer ring at 4~AU. Solids as
large as 10 cm in radius are needed to account for the observed 7 mm
and 3.6 cm flux \citep{Wilner2000ApJ...534L.101W}. The small grains in
the \object{TW~Hya} disk account for 6\% of the total solid mass. We
also estimated a mass in small grains of $3\times 10^{-4}\,M_{\odot}$
assuming that the emission in the millimeter is optically thin, an
average dust temperature of 20~K, and grain opacity $\kappa_\nu = 2.0
(\nu/\nu_0)^{-\beta}$\,g\,cm$^{-2}$ where $\nu_0=230.769$\,GHz and
$\beta=0.6$ \citep{Beckwith1991ApJ...381..250B}.  The two estimates of
dust mass (with radius $a<$~1mm) are consistent within a factor 2 with
each other. The visibility amplitudes generated by the models are
consistent with the observed amplitudes at 345 GHz by
\citep{Qi2004ApJ...616L..11Q}.

For the line observations we augmented the {\it Herschel} data with
{\it SMA} CO 3--2 \citep{Qi2004ApJ...616L..11Q} and {\it JCMT}
$^{13}$CO 3--2 observations \citep{Thi2004A&A...425..955T}.  Following
the characterization of the disk structure from the SED, we ran three
series of models with the thermo-chemical code {\it ProDiMo} \citep[a
detailed description is given in][]{Woitke2009A&A...501..383W}. In
{\it ProDiMo} species abundances are computed at steady-state from the
gas, and dust temperature as well as the local UV field for the
photodissociation reactions. A constant isotopologue ratio
[$^{13}$CO]/[$^{12}$CO] of 69 is assumed. The gas kinetic temperature
is computed by balancing heating and cooling processes. Line profiles
are computed by non-LTE radiative transfer within {\it ProDiMo}. The
disk is assumed to be passively heated. The disk turbulent velocity
and inclination are well constrained by millimeter interferometric
data \citep{Qi2004ApJ...616L..11Q}. The outer disk is irradiated by
direct and scattered stellar photons as well as by interstellar UV
photons. The free parameters of the gas simulations are the disk gas
mass $M_{\mathrm{gas}}$ (between 3 $\times$ 10$^{-4}$ and 0.3
M$_{\odot}$), the fraction of polycyclic aromatic hydrocarbons (PAHs)
in the disk with respect to the interstellar abundance
$f_{\mathrm{PAH}}$, and the cosmic ray flux $\zeta$ (=1.7 $\times$
10$^{-17}$ s$^{-1}$ in the ISM). Observations show that PAHs are
depleted by at least a factor of 10 ($f_{\mathrm{PAH}}=0.1$) in disks
with respect to the interstellar abundance
\citep{Geers2006A&A...459..545G}. Because the gas is mostly heated by
photoelectrons ejected from PAH, the PAH abundance is the main free
parameter that controls the gas temperature.
The three series of models correspond to three possible states: disks
with a very low PAH abundance ($f_{\mathrm{PAH}}=0.01$), disks with a
typical PAH abundance ($f_{\mathrm{PAH}}=0.1$), and X-ray irradiated
disks with a low PAH abundance ($f_{\mathrm{PAH}}=0.01$) but ten times
the standard cosmic ray flux ($\zeta$=1.7 $\times$ 10$^{-17}$
s$^{-1}$) to mimic the influence of strong X-ray emission
\citep{Bruderer2009ApJS..183..179B}. The model results are plotted in
Fig.~\ref{fig_twhya_model}. The density, dust and gas temperature
structure are shown for a typical disk in the appendix. The results
from series 3 are within 10\% of the values of series 2, suggesting
that X-ray does not influence the line fluxes that are emitted at
radii beyond a few AU. In panels a and b we can see that the
OI~63~$\mu$m and 145~$\mu$m flux increases with the disk gas mass. The
OI~63~$\mu$m line is optically thick while the OI~145~$\mu$m line is
optically thin for all models. Both lines arise mostly in a ring
between 4 and 10~AU and thus probe the gas mass up to 10-20AU with
10--20\% contribution from the inner cavity (panel f). In panel c the
CII flux first starts to increase with higher disk gas mass but then
plummets for disk gas masses greater than 10$^{-2}$ M$_{\odot}$. The
CII line is optically thin and the flux increases with radius. As the
disk becomes more massive, more carbon is converted into CO and the
disk becomes cooler. Ionized carbon is excited in gas at
$\sim$~100~K. The CO 3--2 flux increases with increasing disk gas mass
although the emission line is highly optically thick with $\tau>$100
(panel d). CO 3-2 emission comes from the outer disk
($R>$~50~AU). Finally, panel e illustrates the use of the line
emission ratio between two isotopologues ($^{12}$CO and $^{13}$CO) to
constrain column densities or masses. The flux difference between the
two isotopologues shrinks with increasing disk gas mass. The observed
$^{12}$CO/$^{13}$CO 3-2 ratio is consistent with a very low-mass
disk. The CII and CO lines probe the outer disk mass (panel f). All
together, the observations constrain the disk gas mass between 5
$\times$ 10$^{-4}$ and 5 $\times$ 10$^{-3}$ M$_{\odot}$.
\begin{figure*}[!ht]
\centering

{\includegraphics[height=16cm,angle=90]{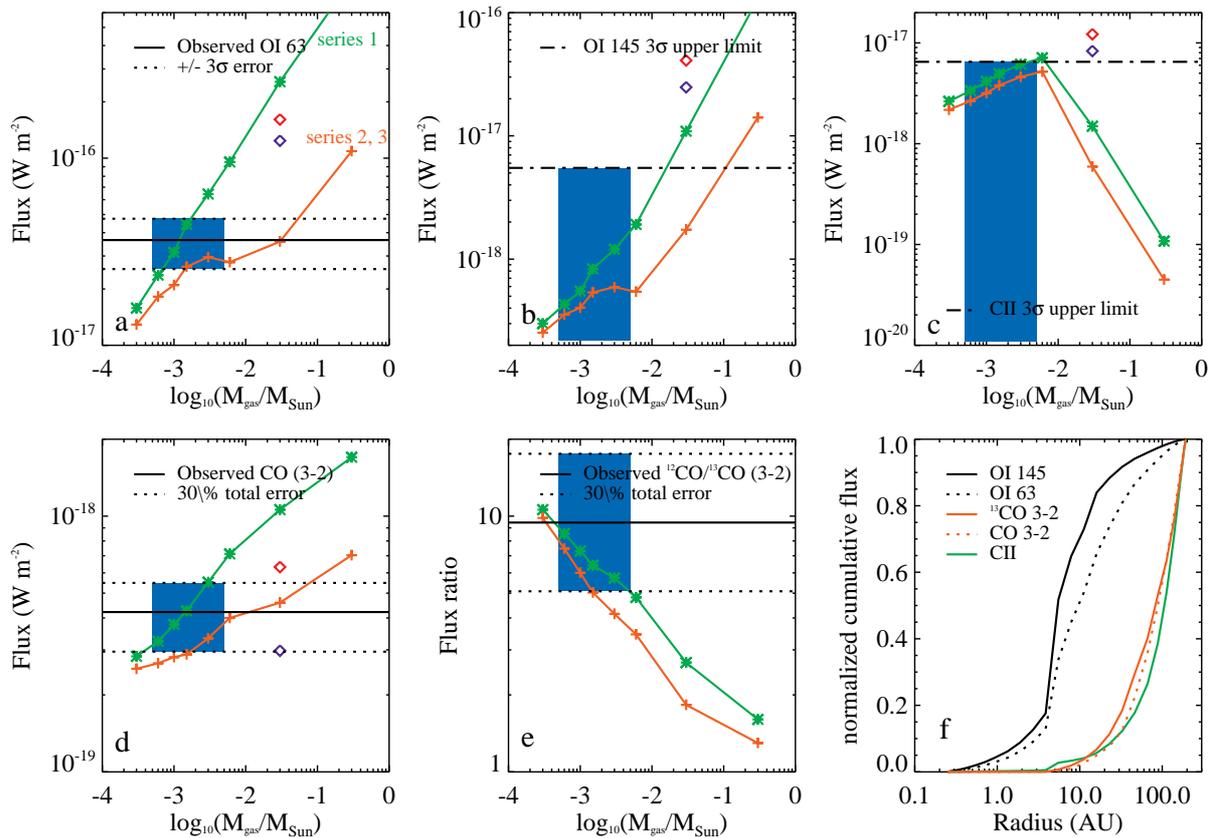}}
\caption{Three series of model results compared to
  observations. The blue boxes enclose the model outputs for disk gas
  mass between 5 $\times$ 10$^{-4}$ M$_{\odot}$ and 5 $\times$
  10$^{-3}$ M$_{\odot}$. Panel a shows the predictions and observation
  of the OI 63~$\mu$m line. The 3$\sigma$ uncertainty range is plotted
  as dashed lines. Panel b and c show the predicted fluxes and the
  3$\sigma$ upper limits for the OI~145~$\mu$m and CII lines. The two
  lower panels (d and e) are the comparison between observations and
  model outputs for $^{12}$CO 3-2 emission and the $^{12}$CO/$^{13}$CO
  3-2 ratio. Panel f shows the normalized cumulative fluxes for a
  10$^{-3}$ M$_{\odot}$ model (series 1).  The diamonds
    (${\color{Red}\diamond}$ $R_{\mathrm{out}}$=174~AU model,
    ${\color{Blue}\diamond}$ $R_{\mathrm{out}}$=120~AU model)
    show the predictions for \object{TW~Hya} from GH08.
    \label{fig_twhya_model} }
\end{figure*}
The modeling suggests that the disk of \object{TW~Hya} has a
gas-to-dust mass ratio of 2.6--26, around a factor of 10 lower than
the interstellar value. If we compare the gas mass to the total mass
in solids (ie including solids with radii up to 10 cm), the
gas-to-solid ratio is 0.17--1.7. \citet[GH08]{Gorti2008ApJ...683..287G}
included X-ray and UV heating in modeling the disk of \object{TW~Hya}
with a gas mass of 0.03 M$_{\odot}$ but noticed that X-ray weakly
influences the fluxes. Their model overestimates the two OI line
fluxes (Table~\ref{table_results} and
Fig.~\ref{fig_twhya_model}). Disk models with X-ray heating only also
predict too strong OI fluxes \citep[M08]{Meijerink2008ApJ...676..518M}
for their model with $L_X$=2 $\times$ 10$^{30}$ erg s$^{-1}$ scaled to
$d$=56~pc.  At $\sim$~10 Myr, \object{TW~Hya} is one of the oldest
classical T~Tauri stars. The outer dust disk is very long-lived, while
the inner disk contains little amount of material. The gas may have a
shorter lifetime than the dust due to photoevaporation or the small
grains result from collisions between the large
grains. \object{TW~Hya} is one of the strongest X-ray active T~Tauri
stars \citep{Raassen2009A&A...505..755R}, which may result in a high
gas photoevaporation rate \citep{Owen2010MNRAS.401.1415O} as evidenced
in the blueshifted [\ion{Ne}{II}] emission observed by
\citet{Pascucci2009ApJ...702..724P}.

\section{Conclusion}

The {\it Herschel-PACS} spectral observations were used to constrain
the gas disk mass surrounding the 10 Myr T~Tauri star
\object{TW~Hya}. We estimate the gas mass to be (0.5--5) $\times$
10$^{-3}$ M$_\odot$ compared to the dust mass
($a_{\mathrm{max}}<$~1mm) of 1.9 $\times$ 10$^{-4}$ M$_\odot$. The
gas-to-dust mass ratio is $\sim$2.6--26, lower than the standard
interstellar value of 100. The ratio gas-to-total-mass in solids is
$\sim$0.17--1.7.  Although the disk is still massive, a significant
fraction of the primordial gas has already disappeared. A large
fraction of the primordial gas may have been evaporated due to the
strong X-ray flux from \object{TW~Hya}. \object{TW~Hya} is the first
example where the disk gas mass around a transitional T~Tauri star can
be determined accurately and directly from gas phase lines. However,
more detailed modeling that includes X-ray physics and $^{13}$CO
photochemistry is needed to confirm the low gas mass.

\begin{acknowledgements}
  W.-F. Thi acknowledges a SUPA astrobiology fellowship.  G. Meeus,
  C. Eiroa, J. Maldonado and B. Montesinos are partly supported by
  Spanish grant AYA 2008-01727. C. Pinte acknowledges the funding from
  the EC 7$^{th}$ Framework Program as a Marie Curie Intra-European
  Fellow (PIEF-GA-2008-220891). D.R. Ardila, S.D. Brittain,
  C.A. Grady, I. Pascucci, B. Riaz, G. Sandell and C. D. Howards,
  J.-P. Williams, G. Matthews, A. Roberge, W. Danchi acknowledge
  NASA/JPL for funding support. E. Solano and J.M. Alacid acknowledge
  the funding from the Spanish MICINN through grant
  AYA2008-02156. The LAOG group acknowledges PNPS, CNES and ANR
  (contract ANR-07-BLAN-0221) for financial support.

\end{acknowledgements}
\bibliographystyle{aa}
\bibliography{14578}

\begin{appendix}

  \section{Density and temperature structure}

\begin{figure}[!ht]
\centering
\resizebox{\hsize}{!}{\includegraphics[]{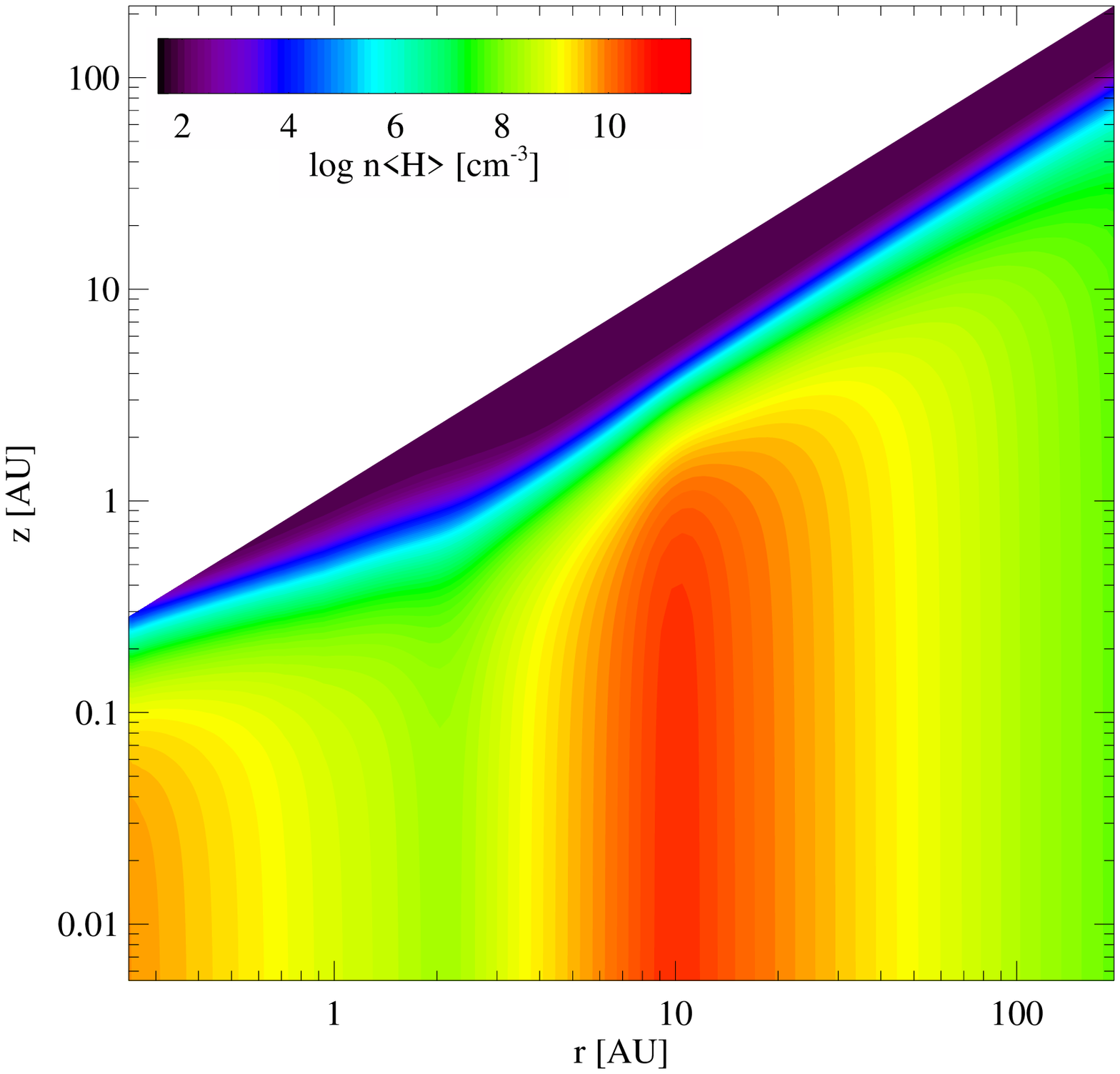}}
\caption{Density profile \label{fig_density_profile}.}
\end{figure}

\begin{figure}
\centering
\resizebox{\hsize}{!}{\includegraphics[]{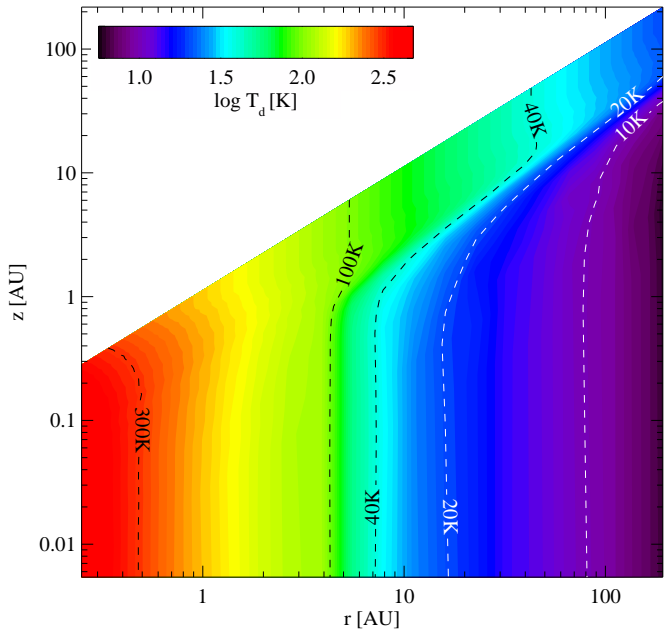}}
\caption{Dust temperature profile. \label{fig_tdust_profile}}
\end{figure}

\begin{figure}
\centering
\resizebox{\hsize}{!}{\includegraphics[]{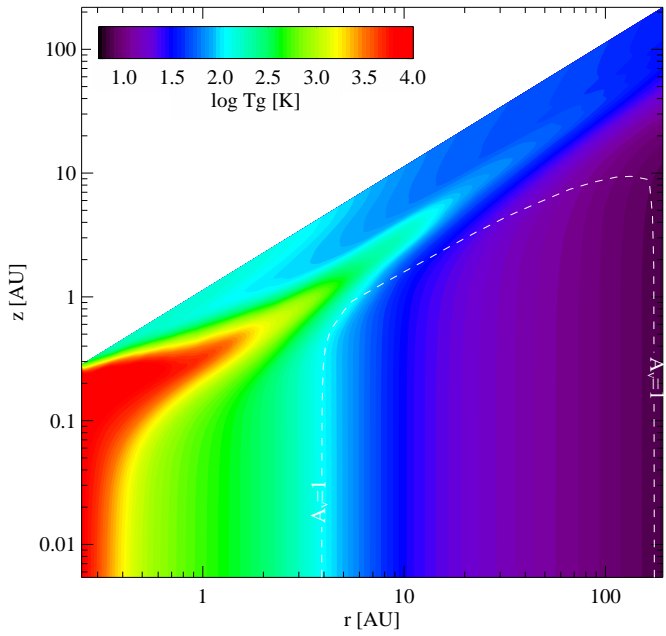}}
\caption{Gas temperature profile. The contour of $A_V$=1 is shown in
  white. \label{fig_tgas_profile}}
\end{figure}

  We show in Fig.~\ref{fig_density_profile},~\ref{fig_tdust_profile},
  and ~\ref{fig_tgas_profile} the density, dust temeprature , and gas
  temperature profile respectively for a disk model with
  $M_{\mathrm{gas}}$=2.9 $\times$ 10$^{-3}$ M$_\odot$,
  $f_{\mathrm{PAH}}$=0.1, and $\zeta$=1.7 $\times$ 10$^{-17}$
  s$^{-1}$. All other parameters are given in
  Table~\ref{disk_parameters}.

\end{appendix}

\end{document}